\documentclass{aa}
\input{epsf.sty}
\usepackage{graphicx}
\usepackage{natbib}

\def\beq{\begin{equation}}
\def\enq{\end{equation}}
\def\ms{${\rm ~M}_{\odot}$}

\title{Discovery of a magnetic neutron star in X-ray transient IGR J01583+6713}

\institute{National Astronomical Observatories, Chinese Academy of
Sciences, Beijing 100012, China}

\author{Wei Wang}

\offprints{W. Wang, wangwei@bao.ac.cn}
\date{Received }

\authorrunning{W. Wang}
\titlerunning{A magnetic neutron star in IGR J01583+6713}

\begin{document}

\abstract {} {IGR J01583+6713 is a new transient source discovered
by the INTEGRAL/IBIS hard X-ray surveys. Optical observations have
provided evidence that it is a high mass X-ray binary, but its
nature remains unclear.} {We used the INTEGRAL/IBIS data to study
IGR J01583+6713 during its outburst.} {We present the temporal
profiles and spectral properties of IGR J01583+6713 around its
outburst on 2005 Dec 6. During the outburst, the mean X-ray
luminosity reached around $4\times 10^{35}$ erg s$^{-1}$ in the
energy range of 20 -- 100 keV. The continuum spectrum can be
fitted by a bremsstrahlung model of $kT\sim 35$ keV or a power-law
model of $\Gamma\sim 2.1$. In addition, the electron resonant
cyclotron absorption lines were detected at $\sim 35$ keV and
possible at $\sim 67$ keV, implying that a magnetic neutron star
of $B\sim 4\times 10^{12}$ G is located in IGR J01583+6713. We
class IGR J01583+6713 as a transient X-ray sources with a magnetic
neutron star. After the outburst, the flux of IGR J01583+6713
decreased and could not be detected by IBIS after 2005 Dec 10.
During the outburst, we cannot confirm the possible pulse period
at 469 s previously reported and did not detect the modulation
signals in the range of 200 -- 2000 s with the IBIS observations.
} {The transient hard X-ray source IGR J01583+6713 contains a
magnetic neutron star, which would help to understand its
transient nature. }

\maketitle

\section{INTRODUCTION}

The transient X-ray source IGR J01583+6713 was discovered by the
the soft Gamma-ray Imager (IBIS/ISGRI) onboard The INTErnational
Gamma-Ray Astrophysics Laboratory (INTEGRAL) during the
observation of Cas A region on 2005 December 6 (Steiner et al.
2005). Subsequent INTEGRAL observations in the same region from
2005 Dec 8 -- 10 showed that the X-ray flux of IGR J01583+6713 was
decreasing (Steiner et al. 2005). Swift observations also found
this source on 2005 Dec 13, and revealed that it is highly
absorbed with $N_{\rm H}\sim 10^{23}$ cm$^{-2}$ (Kennea et al.
2005). Follow-up optical and infrared observations identified the
X-ray source as a Be star (Halpern et al. 2005). Based on optical
spectroscopy, Masetti et al. (2006) classified the counterpart as
an early-type star (O8 III or O9 V) in the Galaxy with a distance
of $\sim 6.4$ kpc, ruling out the possibility of both a supergiant
companion and the source either being a low-mass X-ray binary or a
cataclysmic variable.

Kaur et al. (2008) carried out a multiwavelength study on the
transient source IGR J01583+6713, identified the spectral type of
the companion star to be B2 IVe, and suggested a distance of $\sim
4$ kpc. From the Swift observations, they also reported a possible
pulse period of $\sim 469$ s, but the evidence of pulsation was
only marginal (Kaur et al. 2008), requiring additional
confirmation.

We present results of the INTEGRAL/IBIS observations during the
outburst state of IGR J01583+6713 that occurred in 2005 December.
By reanalyzing these data, we attempt aim to study the hard X-ray
spectral properties of the source both during and after the
outburst. During the outburst state, we attempt search for the
possible pulsation period using IBIS data. In Section 2,
observations and data analysis are presented. We show the spectral
results in Sect. 3 and timing analysis in Sect. 4. A summary of
our results and brief discussions is presented in the last
section.

\section{INTEGRAL/IBIS observations and data analysis}


\begin{figure}
\centering
\includegraphics[angle=0,width=9cm]{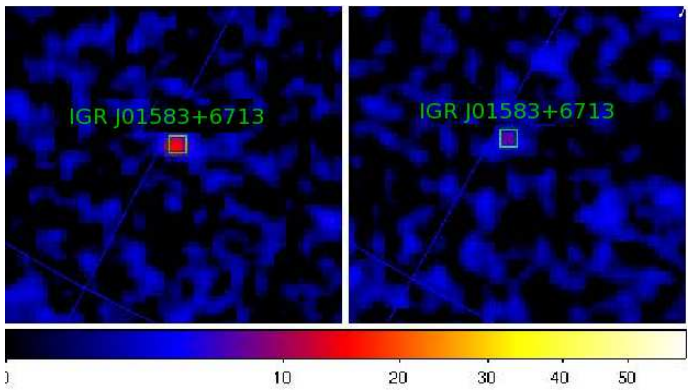}
\caption{Significance mosaic maps around IGR J01583+6713 in
Equatorial J2000 coordinates as seen with INTEGRAL/IBIS in the
energy range of 20 - 60 keV in two revolutions: 384 (left,
outburst), 385 (right, flux decreasing). False color
representation of significance is displayed on a logarithmic
scale. }
\end{figure}

The present database was obtained from the INTEGRAL observations
of the Cas A region performed from 2005 December 5 -- 16 (MJD
53709 -- 53720), from the part of INTEGRAL satellite revolutions
384 -- 387. We used data from the coded mask imager IBIS/ISGRI
(Lebrun et al. 2003; Ubertini et al. 2003) for a total exposure of
$\sim 450$ ks (also see Table 1) after screening for solar-flare
events and erratic count fluctuations due to passages through the
Earth's radiation belts.

The analysis was performed with the standard INTEGRAL off-line
scientific analysis (OSA, Goldwurn et al. 2003) software, ver.
7.0. Individual pointings processed with OSA 7.0 were mosaicked to
create sky images according to the methods and processes described
in Bird et al (2007). And we used the 20 -- 60 keV band for source
detection and to quote fluxes (see Fig. 1 and Table 1).

In revolution 384 (2005 Dec 5 -- 7), X-ray source IGR J01583+6713
was detected at the outburst state with a significance level of
$\sim 16.2\sigma$. And from Dec 8 -- 10, the mean flux of IGR
J01583+6713 decreased to less than half of its value in its
outburst state, with a detection significance level of $\sim
7.3\sigma$. But after 2005 Dec 11, IGR J01583+6713 could not be
detected by IBIS (Table 1). In the present analysis, we use the
data in two revolutions 384 and 385 for the scientific studies.

In the IBIS observations of IGR J01583+6713, it should also be
noted that some observations had large off-axis angles (e.g.,
$>12^\circ$). Because the mask transmission at large off-axis
angles is affected by the mask support structure in a complex way
that cannot not be easily modelled, timing analysis for large
off-axis angles $>12^\circ$ could be subject to significant
systematic errors due to limitations in the off-axis corrections
performed (private communications from the IBIS group). The flux
error bars may also become significantly larger for large off-axis
angles due to lower effective exposures. Therefore, in the
scientific studies (spectral and timing analysis) of IGR
J01583+6713, we only use the data with the off-axis angles
$<12^\circ$ in the following analysis. The remaining on-source
time in the following science analysis is then about 51 ks and 75
ks for Rev 384 and Rev 385, respectively.

Extraction of spectra and light curves for then two revolutions
384 and 385 was also carried out in the OSA software system. In
the timing analysis, because of the sensitivity limits of
IBIS/ISGRI detector, the minimum time intervals are taken to be
about 70 s and $\sim 300$ s for the outburst state Rev 384 and Rev
385, respectively.

\begin{table*}

\caption{INTEGRAL/IBIS observations of the field around IGR
J01583+6713.  }

\begin{center}

\begin{tabular}{l c c c l}

\hline \hline Rev. Num. & Date  & On-source time (ks) & Mean count rate s $^{-1}$ & Detection level \\
\hline 384 & 2005 Dec 05 -- 07 & 118  &  $2.91\pm 0.15$ & 16.2$\sigma$ \\
385 & 2005 Dec 08 -- 10 & 132 & $1.4\pm 0.2$ & 7.3$\sigma$\\
386 & 2005 Dec 11 -- 13 & 121 & $<0.8$ (2$\sigma$) & $<4\sigma$ \\
387 & 2005 Dec 14 -- 16 & 98 & $<0.9$ ($2\sigma$) & $<4\sigma$ \\
\hline
\end{tabular}
\end{center}
{\scriptsize Note - The time intervals of observations in the
revolution number and the corresponding dates, the corrected
on-source exposure time are listed. And mean count rate and the
detection significance level value in the energy range of 20 -- 60
keV were also shown. }
\end{table*}

\section{Spectral properties}

\begin{table*}

\caption{Different fits to the spectrum of IGR J01583+6713 during
the outburst state. }

\begin{center}
\scriptsize
\begin{tabular}{l c c c c l}

\hline \hline Model & $\Gamma$ / $kT$  &  $E_{\rm cyc,0}$ (keV) & $E_{\rm cyc,1}$ (keV) & Flux ($10^{-10}$ erg cm$^{-2}$ s$^{-1}$) & reduced $\chi^2$ \\
\hline
power+cyclabs & $2.13\pm 0.53$ & 35.1$\pm 2.1$ & 65.8$\pm 4.9$ & $1.99\pm 0.42$ & $0.989 (5\ d.o.f)$ \\
bremss+cyclabs & $35.2\pm 4.1$ keV & $35.3\pm 1.6$ & $67.9\pm 4.8$ & $1.98\pm 0.41$ & $0.828 (5\ d.o.f.)$ \\
\hline
\end{tabular}
\end{center}
{\scriptsize Note - {\em power} implies a power law model, {\em
bremss} implies a thermal bremsstrahlung model, {\rm cyclabs}
implies a resonant cyclotron absorption model. The hard X-ray
fluxes in the range of 20 -- 100 keV in different fits are also
given.}

\end{table*}

As a transient X-ray source, IGR J01583+6713 was only detected by
IBIS during the outburst state around 2005 Dec 6 and just few days
after the outburst. In this section, we present the spectral
properties of the transient source IGR J01583+6713 in the range 20
-- 200 keV during the outburst (Rev 384) and after the outburst
(Rev 385) separately. The spectral analysis software package used
is XSPEC 12.4.0x
\footnote{http://heasarc.gsfc.nasa.gov/docs/xanadu/xspec/} (Arnaud
1996).

\begin{figure}
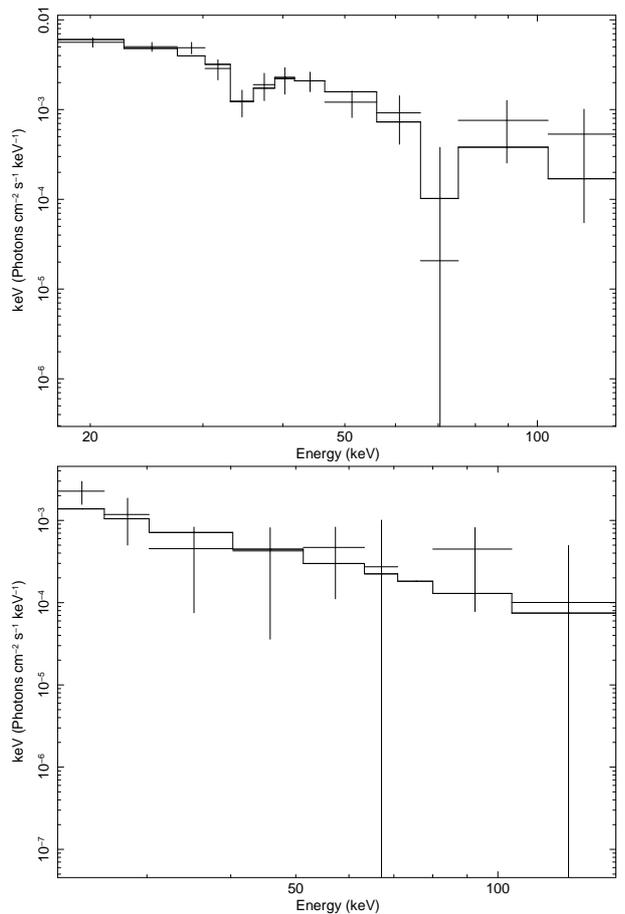

\centering
\includegraphics[angle=-90,width=8cm]{igr0158_spec_384_final.ps}
\includegraphics[angle=-90,width=8cm]{igr0158_spec_385_final.ps}
\caption{{\bf Top:} The hard X-ray unfolded spectrum of IGR
J01583+6713 during the outburst from 2005 Dec 6 --7. The spectrum
was fitted using either a bremsstrahlung model of $kT\sim 35.2\pm
4.1$ plus two cyclotron resonant absorption lines at $\sim 35.3\pm
1.6$ keV and $\sim 67.9\pm 4.8$ keV, or a single power-law model
plus two lines at 35.1$\pm 2.1$ keV and 65.8$\pm 4.9$ keV. {\bf
Bottom:} The hard X-ray unfolded spectrum of IGR J01583+6713 after
the outburst from 2005 Dec 8 --10. The spectrum can be fitted by a
single power law model of $\Gamma\sim 2.7\pm 0.6$. See details in
the text. }
\end{figure}

In revolution 384, IGR J01583+6713 was detected at a high
significance level ($>16\sigma$). The spectrum of this source in
the outburst state is presented in Fig. 2, top panel. We first
used two models to fit the spectrum: a thermal bremsstrahlung
model and a power-law model. The 20 -- 120 keV spectrum was fitted
by a thermal bremsstrahlung model with $kT \sim 25.1\pm 1.9$ keV
(reduced $\chi^2 \sim 1.211$, 11 degree of freedom, thereafter
$d.o.f.$). The derived hard X-ray flux from 20 -- 100 keV was
determined to be $(1.91\pm 0.36)\times 10^{-10}$ erg cm$^{-2}$
s$^{-1}$, corresponding to a hard X-ray luminosity of $\sim
(3.8\pm 0.7)\times 10^{35}$ erg s$^{-1}$ when assuming a distance
of 4 kpc (Kaur et al. 2008). This spectrum was also fitted by a
power-law model, with $\Gamma\sim 2.35\pm 0.18$ (reduced $\chi^2
\sim 1.290, 11 d.o.f.$), and the derived flux (20 -- 100 keV) is
about $(2.2\pm 0.5)\times 10^{-10}$ erg cm$^{-2}$ s$^{-1}$,
corresponding to a hard X-ray luminosity of $\sim (4.4\pm
0.9)\times 10^{35}$ erg s$^{-1}$.

In both fits, we found that an absorption feature around 30 -- 40
keV cannot be fitted, and another possible feature around 60 -- 70
keV also exists. These features are probably related to the
electron cyclotron resonance scattering feature (CRSF) in the
transient source. We therefore used the thermal
bremsstrahlung/power-law model to fit the continuum, and added the
cyclotron resonant absorption model by using the XSPEC cyclabs to
the continuum fit (see Fig. 2 top panel). In the power-law fit
case, we derived $\Gamma\sim 2.13\pm 0.53$ in addition to a CRSF
at $E_{\rm cyc}\sim 35.1\pm 2.1$ keV (F-test probability:
$3.9\times 10^{-9}$) with a line width of FWHM $\sim 1.1\pm 0.4$
keV and a depth 1.75$\pm 0.59$, and another feature at 65.8$\pm
4.9$ keV (F-test probability: 0.009) with a line width of FWHM
$\sim 4.1\pm 2.4$ keV and a depth 0.81$\pm 0.56$ (reduced $\chi^2
\sim 0.989, 5 d.o.f.$). We also carried out the fit with a
bremsstrahlung model plus the CRSF on the spectrum. We then also
found two absorption features in the fit, and obtained $kT\sim
35.2\pm 4.1$ keV plus a CRSF at $35.3\pm 1.6$ keV (F-test
probability: $5.6\times 10^{-10}$) with a FWHM $\sim 1.2\pm 0.4$
keV and a depth of $\sim 1.98\pm 0.71$, the other CRSF at $67.9\pm
4.8$ keV (F-test probability: 0.011) with a FWHM $\sim 4.4\pm 2.5$
keV and a depth of $\sim 0.93\pm 0.55$ (reduced $\chi^2 \sim
0.828, 5\ d.o.f.$, see Fig. 2, top panel \& Table 2).

After the outburst from 2005 Dec 8 -- 10, IGR J01583+6713 was
detected by IBIS at a lower significance level $\sim 7\sigma$. The
spectrum after the outburst is displayed in the bottom panel of
Fig. 2. The 20 -- 100 keV spectrum could be described by a single
power-law model of the photon index $\Gamma\sim 2.7\pm 0.6$
(reduced $\chi^2 \sim 0.929, 7 d.o.f$). No significant absorption
features are detected. The derived flux is $\sim (7.1\pm
2.8)\times 10^{-11}$ erg cm$^{-2}$ s$^{-1}$ from 20 -- 100 keV,
corresponding to a mean luminosity of $\sim (1.4\pm 0.5)\times
10^{35}$ erg s$^{-1}$.

\section{Timing analysis}

From one SWIFT observation, during which the X-ray intensity was
higher, Kaur et al. (2008) had a possible pulse detection with a
period of 469.2 s. This detection was, however marginal. Thus,
using the data of the IBIS observations during the outburst state
of the transient hard X-ray source IGR J01583+6713, we attempt to
search for the possible pulse period.

\begin{figure}
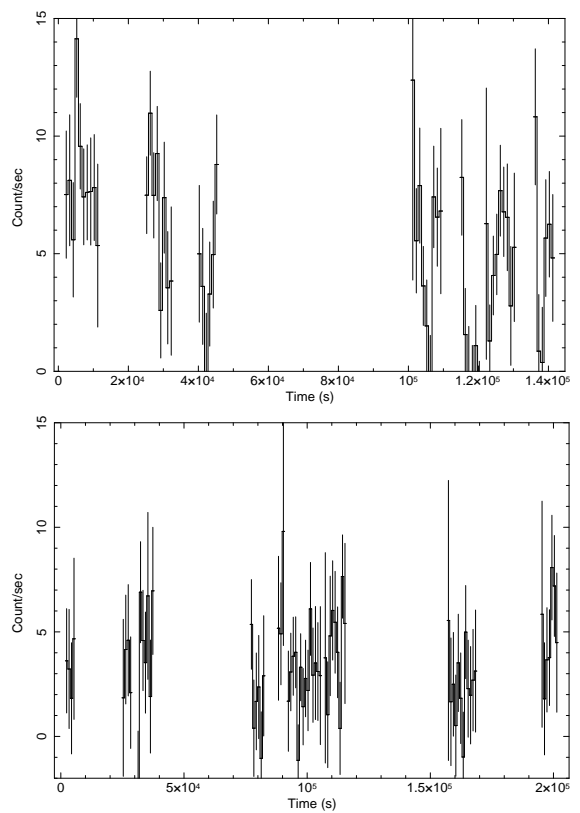

\centering
\includegraphics[angle=-90,width=7.5cm]{igr0158_384_lc_final.ps}
\includegraphics[angle=-90,width=7.5cm]{igr0158_385_lc_final.ps}
\caption{The IBIS/ISGRI background subtracted light curve of IGR
J01583+6713 from 20 -- 60 keV. {\bf Top:} The light curve in Rev
384 starting at MJD 53710.101 (2005 Dec 6) with the bin time of
1000 s. {\bf Bottom:} The light curve in Rev 385 starting at MJD
53712.159 (2005 Dec 8) with the bin time of 1000 s.}
\end{figure}

\begin{figure}
\centering
\includegraphics[angle=0,width=8cm]{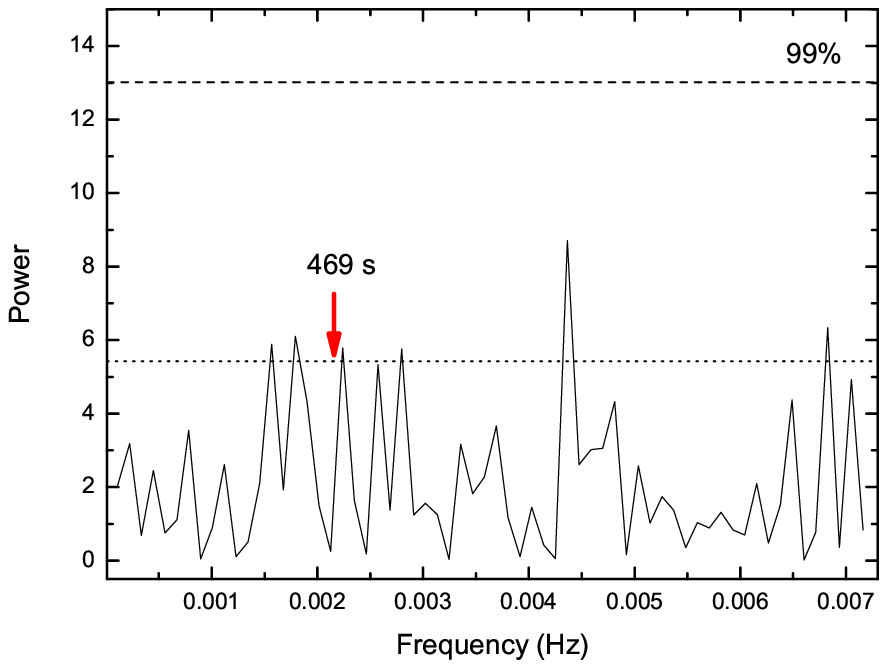}
\includegraphics[angle=0,width=8cm]{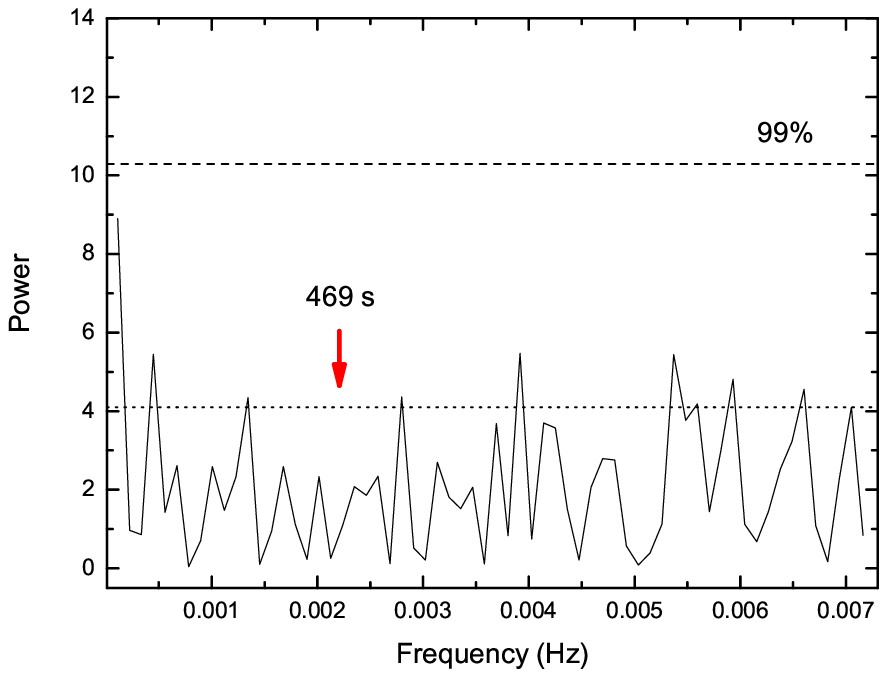}
\caption{The power spectra of the IBIS light curve of IGR
J01583+6713 for the outburst state on 2005 Dec 6: (top) from MJD
53710.101 -- 53710.219; (bottom) MJD 53710.370 -- 53710.457. The
possible period at 469 s reported by Kaur et al. (2008) is also
noted in both power spectra. The dashed and dotted lines present
the $99\%$ significant levels for white noise and red noise
respectively. }
\end{figure}

The light curve of IGR J01583+6713 from 2005 Dec 6 -- 7 observed
by IBIS/ISGRI in the energy range of 20 -- 60 keV is shown in Fig.
3 top panel. The background count rate was subtracted and the
barycentric corrections were also performed. The count rates were
rebinned in intervals of 1000 s in the plot for clarity. During
the outburst on 2005 Dec 6, the count rate reached $\sim 10$ cts
s$^{-1}$. On Dec 7, the rate decreased to $\sim 5$ cts s$^{-1}$.
From 2005 Dec 8 -- 10 (revolution 385, Fig. 3 bottom panel), the
count rate was around 3 cts s$^{-1}$. After the outburst on Dec 6,
the X-ray flux of the transient source IGR J01583+6713 decreased.

\begin{figure}
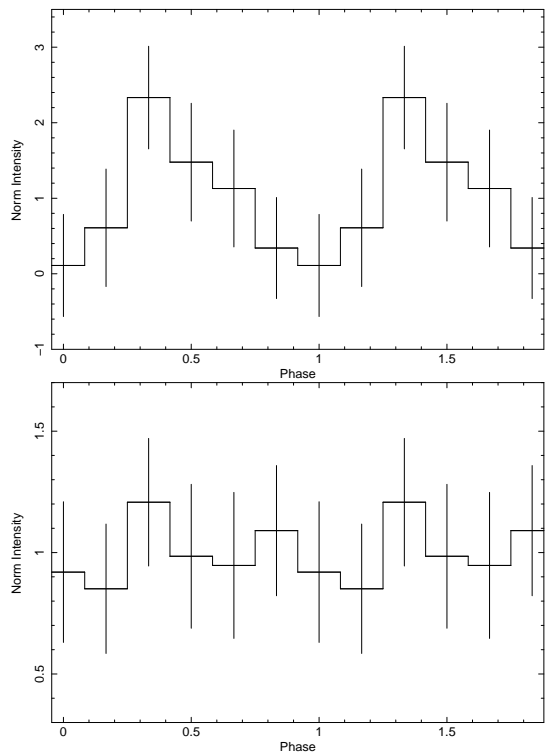

\centering
\includegraphics[angle=-90,width=7cm]{igr0158_fold_b1_469.ps}
\includegraphics[angle=-90,width=7cm]{igr0158_fold_b2_469.ps}
\caption{The IBIS/ISGRI light curves (20 -- 60 keV) of IGR
J01583+6713 folded at a possible pulsation period (469 s) for the
observational data of two time intervals in outbursts: MJD
53710.101 -- 53710.219 (top); and MJD 53710.370 -- 53710.457
(bottom). The pulse profile is repeated once for clarity. }
\end{figure}

\begin{figure}
\centering
\includegraphics[angle=-90,width=7cm]{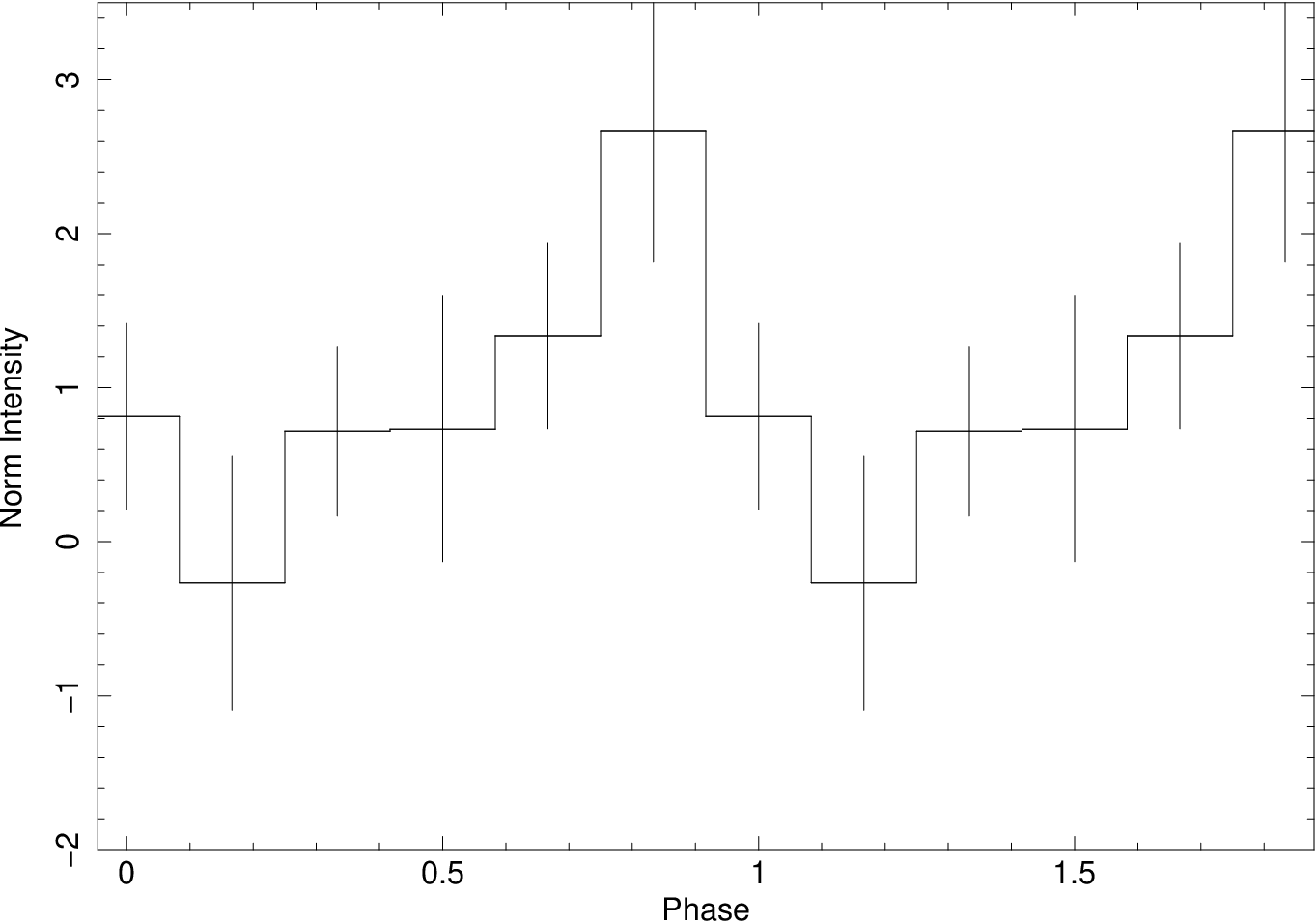}
\includegraphics[angle=-90,width=7cm]{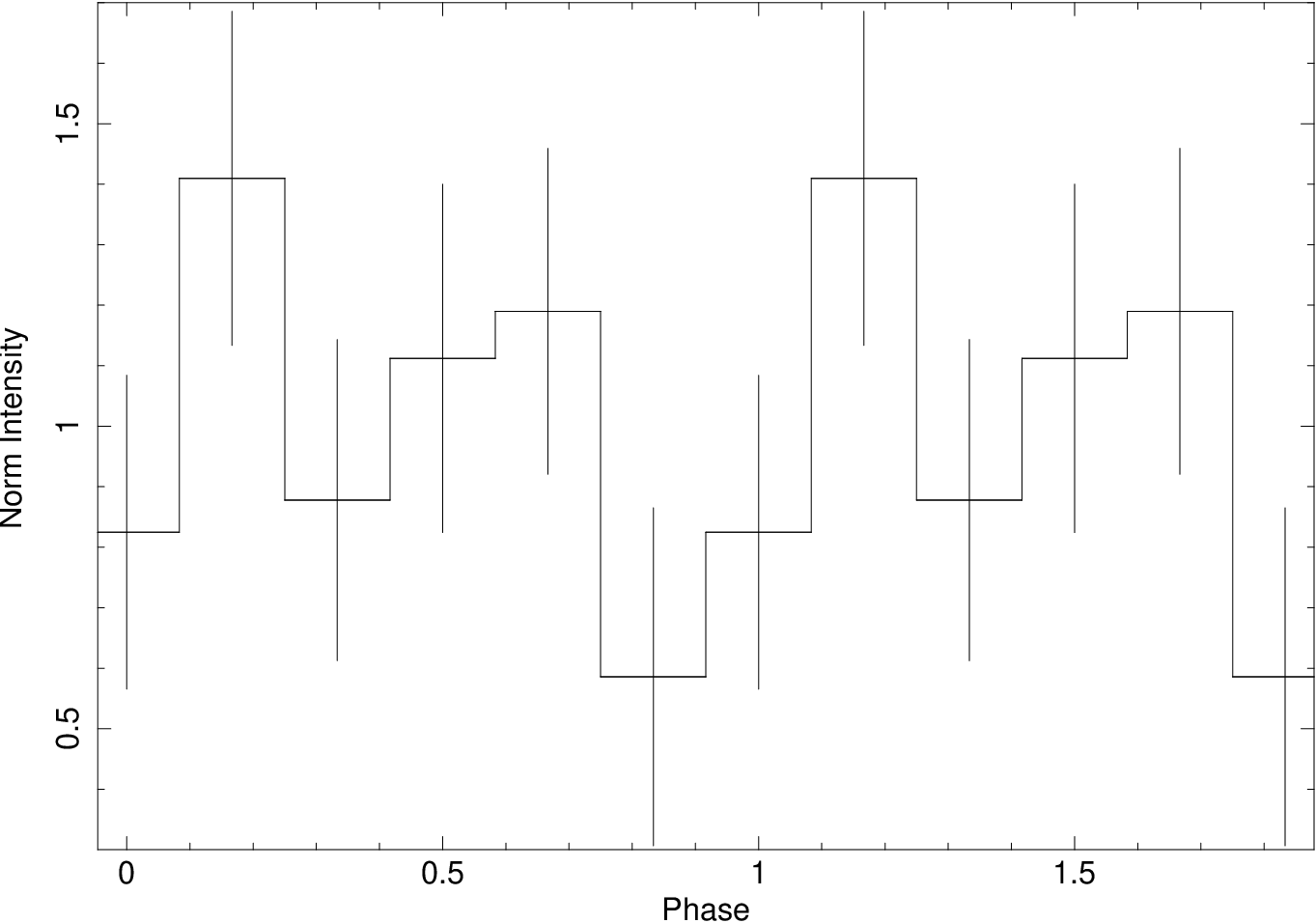}
\caption{The IBIS/ISGRI light curves (20 -- 60 keV) of IGR
J01583+6713 folded at a possible pulsation period (230 s) for the
observational data of two time intervals in outbursts: MJD
53710.101 -- 53710.219 (top); and MJD 53710.370 -- 53710.457
(bottom). The pulse profile is repeated once for clarity. }
\end{figure}

To search for the modulation signal in the X-ray light curves, we
applied the FFT to the IBIS/ISGRI light curves of IGR J01583+6713
during the outburst on 2005 Dec 6. We used {\em powspec} of the
{\em Xronos} software package ver. 5.21
\footnote{http://heasarc.gsfc.nasa.gov/docs/xanadu/xronos/}
available from NASA's HEASARC for the power density spectrum
analysis. To estimate the significance level of the possible
detected peak signals in the power spectra, we considered the
white and red noise significance levels. The $99\%$ white noise
significance level was estimated using Monte Carlo simulations
(see examples in Kong et al. 1998). The $99\%$ red noise
significance level was estimated using the REDFIT35 subroutine
(Schulz \& Mudelsee 2002)
\footnote{ftp://ftp.rz.uni-kiel.de/pub/sfb313/mschulz/redfit35.zip},
which fits a first-order autoregressive process to the time series
to estimate the red-noise spectrum.

\begin{figure}
\centering
\includegraphics[angle=0,width=9cm]{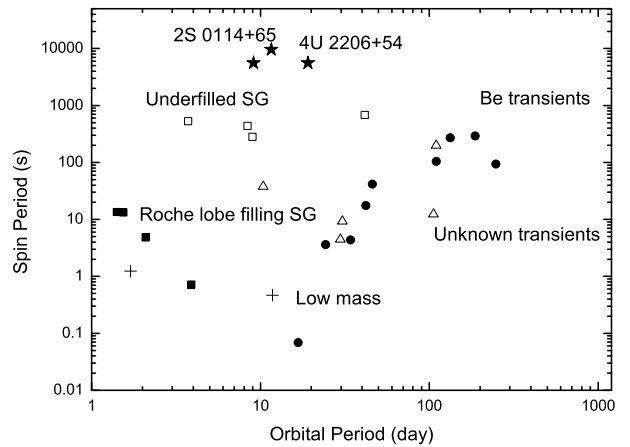}
\caption{The spin period - orbital period diagram for the
accreting neutron star systems (the Corbet diagram, Corbet 1986).
The data point for 2S 0114+65 is taken from Farrell et al. (2008);
the data point for 4U 2206+54 is taken from Wang (2009), two
possible values of orbital period: 9.6 day and 19.1 day are both
shown in this diagram; other points are taken from Bildsten et al.
(1997). There exist a positive correlation between $P_{\rm spin}-
P_{\rm orbit}$ for the Be transient systems; and a possible
negative correlation for the disk-accretion system including Roche
lobe filling supergiants and low mass systems. For the underfilled
Roche lobe supergiants and two long-pulsation systems 2S 0014+65
and 4U 2206+54, which all belong to wind-fed accretion systems,
the possible relation $P_{\rm spin}- P_{\rm orbit}$ differing from
that of the Be transients. }
\end{figure}

The hard X-ray count rates of IGR J01583+6713 in the outburst were
obtained in several time intervals (see Fig. 3 top panel). In two
time intervals, MJD 53710.101 -- 53710.219 and MJD 53710.370 --
53710.457, the detected flux from IGR J01583+6713 had a count rate
higher than 5 cts s$^{-1}$, so we applied the FFT to these two
light curves to search for possible modulation features. The
derived power spectra of two outburst light curves were presented
in Fig. 4, binned in 70 s intervals. The position of the pulse
period at 469 s was identified in two power spectra, but no
significant peak signals were found around 469 s in both power
spectra. For the time interval MJD 53710.101 -- 53710.219, we
found a possible peak feature at $\sim 230$ s (Fig. 4 top panel),
but it may be caused by the noise (significantly lower than the
$99\%$ white noise level and just a little higher than the $99\%$
red noise level). In addition, this possible feature around 230 s
was not found in the time interval MJD 53710.370 -- 53710.457.
Therefore, using the power spectrum analysis of our present IBIS
data, we did not detect any possible modulation period signals in
the X-ray light curves from 200 -- 2000 s.

We note that the 469 s period was found by means of the
pulse-folding technique (Kaur et al. 2008). We were also able to
derive the folded profiles by folding the hard X-ray light curves
at the possible pulse period. We folded the X-ray light curves
using the {\em efold} tool of the {\em Xronos} software package.
The folded pulse profiles were derived for two possible period
values indicated in this work 469 s and 230 s (see Fig. 5 \& 6).
The folded profiles at both two periods of 469 s and 230 s from
MJD 53710.101 -- 53710.219 show the possible pulse features with
the pulse fraction near $100\%$, if we assume this pulse period to
be true. From MJD 53710.370 -- 53710.457, significant pulse
features do not appear in the folded profiles (large error bars
can make the profiles look almost flat). When only considering
pulse-folded profiles, we may find two possible period values: 469
s and 230 s. Only one of them may be the true period or both of
them are not. However, both two period signals have not been
detected in power spectrum analysis. Therefore the pulse-folding
technique would not be a good tool for blindly searching for the
possible period signal at the first step. The possible pulse
period of 469 s cannot be confirmed with the IBIS data.

In addition, we could also check the possible pulse period at 469
s in a different way. In Fig. 7, we show the $P_{\rm spin}-P_{\rm
orbit}$ diagram for known neutron star X-ray binaries (Corbet
diagram, Corbet 1986). From optical identification (Masetti et al.
2006), IGR J01583+6713 may be a Be/X-ray star that might follow
the previously found correlation between orbital period and pulse
period in the classic Be star/neutron star binary systems (Corbet
1986). With the possible pulse period at 469 s, Kaur et al. (2008)
estimated that the X-ray binary IGR J01583+6713 has an orbital
period in the range 216 -- 561 days assuming the maximum
eccentricity of the orbit to be 0.47 for Be binaries as observed
by Bildsten et al. (1997). Some neutron-star binary systems like
underfilled Roche lobe supergiants and 4U 2206+54, 2S 0114+65 are
powered by direct wind accretion. These systems may follow a
$P_{\rm spin}-P_{\rm orbit}$ relation that is quite different from
that of the Be transient systems. If IGR J01583+6713 were to
follow the possible relation (i.e., one that is not clearly
evident) of wind accretion systems, assuming a pulse period of 469
s, IGR J01583+6713 may have a possible orbital period in the range
3 -- 12 days.

To search for the possible orbital modulation signal of IGR
J01583+6713, we used the archival one-day average RXTE/ASM data
\footnote{ASCII files available from http://xte.mit.edu/}. We
downloaded the one-day average RXTE/ASM data for IGR J01583+6713
from 2000 Jan to 2009 Dec and applied the FFT to the ten years of
light curve data. The power spectrum of the RXTE/ASM data is
presented in Fig. 8, but no significant features are apparent. In
the range of 216 -- 561 days, no peak features were found, and in
the range of 3 -- 12 days, two possible signals at $\sim 7.7$ days
and $\sim 10.6$ days could be noise signals which are
significantly lower than the $99\%$ white noise level. Therefore,
with the RXTE/ASM data, we have not found any possible orbital
period within the interval 3 -- 1000 days, so we also cannot
confirm the possible pulsed period at 469 s yet assuming the
possible correlation between the spin period and orbital period in
neutron star binaries.

\begin{figure}
\centering
\includegraphics[angle=0,width=9cm]{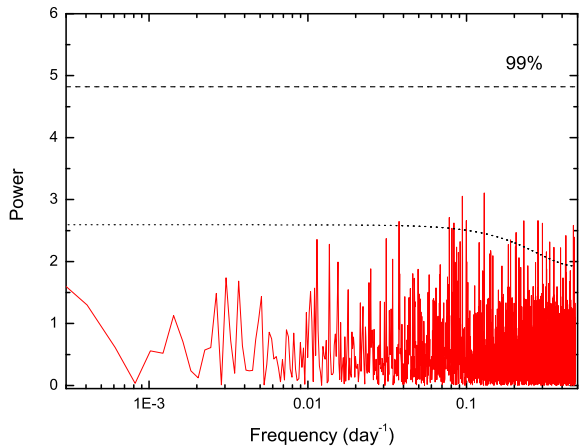}
\caption{The power spectrum of the ASM light curve of IGR
J01583+6713 from 2000 Jan to 2009 Dec. We have not detected any
significant modulation signals during the interval 4 -- 1000 days.
The possible peak features at $\sim 7.7$ days and $\sim 10.6$ days
are probably noise signals, which are significantly lower than the
$99\%$ white noise level (dashed line). The dotted line presents
the $99\%$ red noise significant level. }
\end{figure}

\section{Summary and discussion}

We have attempted to determine the nature of the transient hard
X-ray source IGR J01583+6713 using IBIS/ISGRI observations
acquired during the outburst state around 2005 Dec 6. During the
outburst, IGR J01583+6713 had an average luminosity of $\sim
4\times 10^{35}$ erg s$^{-1}$ in the energy interval 20 -- 100
keV. The spectrum of IGR J01583+6713 can be described by both a
thermal bremsstrahlung model with $kT \sim 35$ keV plus two
resonant cyclotron absorption lines at $\sim 35.3\pm 1.6$ keV and
$\sim 67.9\pm 4.8$ keV and a single power-law model plus two lines
at 35.1$\pm 2.1$ keV and 65.8$\pm 4.9$ keV. We can estimate the
magnetic field of the compact object in IGR J01583+6713 according
to a formula \beq [B/10^{12}{\rm G}]=[E_{\rm cycl}/11.6{\rm
keV}](1+z), \enq where $E_{\rm cycl}$ is the energy of the
fundamental line, which we assume to be $E_{\rm cycl}=35\pm 3$
keV, and $z$ is the gravitational redshift near the surface of the
neutron star. For a canonical neutron star of 1.4 \ms with a
radius of 10 km, we can take $z\sim 0.3$ (Kreykenbohm et al.
2004). Then we found a magnetic field of $\sim (4.0\pm 0.4) \times
10^{12}$ G for IGR J01583+6713. Thus, the compact object in IGR
J01583+6713 should be a magnetic neutron star.

Timing analysis was also carried out during the outburst to search
for the possible pulsation period. The possible pulsation period
at $\sim 469$ s reported by Kaur et al. (2008) cannot be confirmed
by our power spectrum analysis. In the time interval of MJD
53710.101 -- 53710.219, a possible feature at 230 s was found in
the power spectrum, which is still significantly lower than the
$99\%$ white noise level. However, the folded light curves at both
230 s and 469.2 s from MJD 53710.101 -- 53710.219 show possible
pulse profiles. The pulse-folding technique may therefore provide
some incorrect information about the blind period search. In
addition, with the present IBIS observations, we have not detected
any possible modulation period in the range 200 -- 2000 s, and
present RTXE/ASM data does no exhibit any orbital modulation
signals in the range 3 -- 1000 days.

After the outburst, the hard X-ray flux of IGR J01583+6713
decreased according to our observations. During the interval 2005
Dec 8 -- 10, the mean X-ray luminosity was around $\sim 1.4\times
10^{35}$ erg s$^{-1}$ in the range of 20 -- 100 keV. The spectrum
could be fitted by a power-law model of $\Gamma\sim 2.7$. Because
of the detection of a low significance level ($\sim 7\sigma$), the
CRSF was not found after the outburst. The derived power-law slope
of $\sim 2.7$ is generally steeper than those of other neutron
star X-ray binaries (around 1.7 -- 2.1, e.g. see Wang 2009;
Farrell et al. 2008). A large uncertainty of 0.6 still exists,
which requires confirmation.

The transient X-ray pulsar IGR J01583+6713 may represent a link to
the supergiant fast X-ray transients (SFXTs) in some supergiant
high mass X-ray binaries (e.g., Sguera et al. 2006) belonging to
wind-fed systems. These transients show soft gamma-ray
time-structured bursts of durations from several hours to about 2
days detected by INTEGRAL/IBIS observations. The mean X-ray
luminosity reaches around $10^{35}-10^{36}$ erg s$^{-1}$, while in
quiescence, the X-ray luminosity is only $\sim 10^{33}$ erg
s$^{-1}$ (Sguera et al. 2006). These behaviors of SFXTs are
similar to those of the transient X-ray pulsar IGR J01583+6713,
though IGR J01583+6713 has a different companion from SFXTs. Soft
X-ray monitoring of IGR J01583+6713 found that in quiescence after
the outburst, the X-ray luminosity of IGR J01583+6713 decreased to
several $10^{33}$ erg s$^{-1}$ (Kaur et al. 2008). Both the origin
of transient X-ray pulsars and the mechanisms that produce SFXTs
are at present unkown. Connections between them would help us to
understand common physical properties in transient or suddenly
enhanced accretion systems. Thus, future detailed studies of the
transient source IGR J01583+6713 would certainly enhance our
understanding of the physical origin and connections of both SFXTs
and transient X-ray pulsars.

\section*{Acknowledgments}
We are grateful to the referee for the suggestions of improving
the manuscript and to X.L. Zhang for the preparation of some
figures. This paper is based on observations of INTEGRAL, an ESA
project with instrument and science data center funded by ESA
member states (principle investigator countries: Demark, France,
Germany, Italy, Switzerland and Spain), the Czech Republic and
Poland, and with participation of Russia and US. W. Wang is
supported by the National Natural Science Foundation of China
under grants 10803009, 10833003.

\end{document}